# Unravelling The potential of Hybrid Borocarbonitride Biphenylene 2D Network for Thermoelectric Applications: A First Principles Study

Ajay Kumar[a] Parbati Senapati[a] and Prakash Parida[a]*
[a]Department of Physics, Indian Institute of Technology Patna, Bihta, Bihar, India, 801106
*Corresponding author: - pparida@iitp.ac.in

**Abstract**

In this study, we investigate a novel hybrid borocarbonitrides (bpn-BCN) 2D material inspired by recent advances in carbon biphenylene synthesis, using first-principles calculations and semi-classical Boltzmann transport theory. Our analysis confirms the structural stability of bpn-BCN through formation energy, elastic coefficients, phonon dispersion, and molecular dynamics simulations at 300 K and 800 K. The material exhibits an indirect band gap of 0.19 eV (PBE) between the **X** and **Y** points and a direct band gap of 0.58 eV (HSE) at the **X** point. Thermoelectric properties reveal a high Seebeck coefficient, peaking at $700 VK^{-1}$ for n-type carriers at 200K along the x-axis, while n-type has a maximum of $588 VK^{-1}$. The electrical conductivity is $2.2 \times 10^7 \Omega^{-1} m^{-1}$ for hole carriers, surpassing that of conventional 2D materials. The consequences of the high Seebeck coefficient and conductivity reflect a high-power factor with a peak value of $178 \times 10^{-3} Wm^{-1}K^{-2}$ at 1000K for p-type carriers along the y-axis whereas, for n-type $91 \times 10^{-3} Wm^{-1}K^{-2}$. Moreover, the highest observed *zT* values were 0.78 (0.72) along the x (y) direction at 750 K for p-type and 0.57 (0.53) at 750 K along the x (y) axis for n-type. Our findings suggest that the bpn-BCN 2D network holds significant potential for thermoelectric applications due to its exceptional performance.

1. **Introduction**

The growing global demand for effective energy production, conservation, and management has an imperative interest in diverse renewable sources like solar, wind, and biomass[1-3]. Alongside these, thermoelectric (TE) materials offer another option for power generation, harnessing heat from various sources, including fossil fuel combustion, sunlight, or industrial processes[4-6]. Thermoelectric materials primarily generate power from heat energy, which can be sourced from domestic to industrial scales. Energy conservation or waste-heat harvesting involves capturing and converting excess heat into usable electrical power. The extent of their impact depends on their efficiency, yet it is unrealistic to expect thermoelectric energy conversion alone to solve all energy challenges. However, these materials are anticipated to contribute alongside other technologies to address energy efficiency concerns increasingly. Thermoelectric material works on the Seebeck effect, converting temperature gradient into voltage gradient[7]. They have significant potential in waste heat recovery from industrial processes, transforming excess heat into electricity to enhance energy efficiency and reduce environmental impact. Thermoelectric generators are integrated into portable electronic devices, potentially replacing traditional batteries[8,9]. TE material-based clothing could harvest body heat to supplement power, extending device runtime and enhancing user convenience[10,11].

The performance of a TE device depends on two key factors: the temperature gradient (ΔT) and the intrinsic material property known as the thermoelectric figure of merit (*zT*). Further, the efficiency of the TE module for power generation combines the Carnot

efficiency $\left(\frac{\Delta T}{T_H}\right)$ with the figure of merit in the given equation: -

$$\eta = \frac{\Delta T}{T_H} \frac{\sqrt{(1+zT)}-1}{\sqrt{(1+zT)}+T_C/T_H} \qquad (1)$$

Where $\Delta T = T_H - T_C$; $T_H$ and $T_C$ is the temperature of the hot and cold end of the module. **Eq. 1** illustrates that enhancing efficiency necessitates high $zT$ values and a substantial temperature gradient across the thermoelectric materials. The dimensionless figure of merit is defined as

$$zT = \frac{S^2 \sigma T}{\kappa} \qquad (2)$$

where S, σ, and T represent the Seebeck coefficient, electrical conductivity, and absolute temperature, respectively. Here, κ ) denotes the total thermal conductivity comprising electronic $\kappa_e$ and phononic $\kappa_{ph}$ contributions. These parameters are intricately linked to the specifics of the electronic structure and the scattering of charge carriers (electrons or holes), which strongly influence each other. The quantity $S^2\sigma$ is the power factor (PF) that is essential for achieving high performance. A large PF indicates the generation of a high voltage and high current. A high $zT$ value suggests a favourable combination of high PF and low κ, indicating promising thermoelectric properties.

In order to achieve the high $zT$ value, the material with high electronic conductivity and low phononic (lattice) conductivity will be recommended. Slack introduces a phonon-glass electron-crystal (PGEC) concept that aims to optimize thermoelectric performance by selectively scattering phonons while enabling the free movement of electrons[12]. The goal is to scatter phonons effectively to hinder heat flow while allowing electrons to move freely for efficient electrical conductivity. Generally, metallic materials have high σ but also high $\kappa_e$, while insulators lack both, making them unsuitable for TE. Balancing σ, κ, and other properties is crucial. Semiconductors with small bandgaps with lower $\kappa_{ph}$ are preferred for TE applications.

There are a number of ways to manipulate the phononic part of κ. For illustration, introducing defects such as vacancies or isotopes disrupts the ordered structure, leading to phonon scattering and a consequent reduction in $\kappa_{ph}$[13, 14]. Also, complex crystal structures with extended phonon scattering paths hinder heat transfer by scattering of phonons. These structures contain obstacles like impurities and intricate atomic arrangements, which scatter the phonons and dissipate their energy[15, 16]. Nano-dimensional materials offer unique advantages for thermoelectric due to enhanced phonon scattering, reduced thermal conductivity, and increased electrical conductivity at the nanoscale[17-20]. Their tunability enables the customization of band structures and electronic properties for achieving desirable TE characteristics like high Seebeck coefficients and $zT$ values[21]. The twistronics of graphene-based heterostructures have been reported to lower the phonon thermal conductivity.[22] 2D materials offer advantages such as reduced dimensionality, enhanced thermoelectric performance through quantum confinement effects and increased phonon scattering at interfaces. 2D materials like SnSe, $Bi_2Te_3$, and PbTe have been extensively studied theoretically and experimentally due to their potential for high thermoelectric performance[23-28]. Their layered structure allows for efficient manipulation of electronic transport properties and phonon scattering, crucial factors in thermoelectric conversion. For instance, SnSe has found its exceptionally high thermoelectric performance, owing to its intrinsically low thermal and high electrical conductivity. $Bi_2Te_3$, a classic thermoelectric material, has been explored in its 2D form to leverage the advantages of reduced dimensionality, potentially enhancing its thermoelectric efficiency[29]. These traditional TE materials are restricted because of their high cost, toxicity, and scarcity. Therefore, it is urgent to explore new TE materials to overcome these limitations.

Qitang Fan et al. recently synthesised the ultra-flat $sp^2$ hybridized non-benzoide atomic thin 2D network of carbon atoms by a bottom-up approach named biphenylene network (BPN). Scanning tunnelling spectroscopy indicates that the conductivity of the biphenylene network varies with its width[30]. Theoretically, metallic phases are an ideal 2D infinite sheet[31, 32]. On the other hand, the boron and nitrogen networks of biphenylene are reported to have a finite bandgap due to the ionic nature of the B-N bond. A systemic B-N network has a wider bandgap of 3.2 eV[33]. In our previous work, we reported the bandgap of the B-N biphenylene (I-BPN) monolayer of 1.73 eV by introducing the B-B and N-N bonds along with the B-N bond in the biphenylene network structure[18]. It is expected that a similar trend will be seen in the graphene and hexagonal boron nitrogen monolayer with semi-metal and insulator (~5.6 eV), respectively [34]. Researchers have explored biphenylene networks composed of various materials, including planar structures like SiC and AlN, as well as buckled configurations such as Si, Ge, BAs, and ZnSe, providing a range of band gaps, offering diverse electronic properties for potential applications[35]. A small bandgap with low phonon thermal conductivity is necessary for thermoelectric application. Therefore, the hybrid composition of boron-carbon and nitrogen (BCN) in biphenylene 2D sheets will be a better option. The honeycomb structure of BCN had been synthesised and explored their electrocatalytic and catalytic properties of the electrochemical and photochemical processes even better than the conventional catalyst Pt[35, 36]. Apart from experimental investigations, computational studies have delved into the thermal, electrical, and mechanical stability of ternary BCN monolayers, exploring diverse chemical compositions[37-39].

Inspired by Fan et al., this study explores the structural, electronic, and thermoelectric properties of BCN biphenylene (bpn-BCN) 2D networks. Due to its unique composition, the hybrid BCN structure holds significant promise as a TE material. It features partially ionic bonds between B and N, fostering charge localization alongside charge delocalization facilitated by C-C covalent bonds. The stoichiometry of BCN allows for tunable bandgaps, further enhancing its potential. Additionally, it exhibits promise in reducing phonon thermal conductivity through high phonon-phonon scattering resulting from the varying atomic masses of B, C, and N. Despite minor differences in mass, these variations lead to significant phonon scattering, offering a pathway for enhancing thermoelectric performance. Our investigation assesses structural stability via formation energy, phonon dispersion calculations, mechanical strength (Young modulus and Poisson ratio), and ab-initio molecular dynamics (AIMD). Furthermore, we determine lattice thermal properties using a semi-classical transport equation for phonons, including the thermal conductivity, phonon group velocity, and phonon lifetime. Additionally, we analyze the Seebeck coefficient, electrical conductivity, and electronic thermal conductivity as functions of temperature and chemical potential, utilizing semi-classical transport equations for electrons.

## 2. Computational details

The bpn-BCN monolayer has been studied using Density Functional Theory (DFT) with the Vienna Ab Initio Simulation Package (VASP)[40]. This study utilizes the Projector Augmented Wave (PAW) method, essential for accurately modelling interactions between valence and core electrons, along with periodic boundary conditions. Pseudopotentials for B, C, and N were derived from their respective electronic configurations: $2s^2 2p^1$, $2s^2 2p^2$, and $2s^2 2p^3$. The exchange-correlation potential was treated with the Generalized Gradient Approximation (GGA) formulated by Perdew–Burke–Ernzerhof (PBE). This GGA-PBE functional is commonly used for 2D materials, particularly those composed of low atomic number elements, due to its effective balance between computational accuracy and cost without significantly compromising qualitative results. However, GGA-PBE is known sometimes to underestimate the electronic bandgap. Therefore, we employed the Heyd–Scuseria–Ernzerhof[39] (HSE) functional to obtain more accurate bandgap predictions, which provide more precise results, although more computationally demanding. All the electronic and thermoelectric calculations have been performed using the HSE functional.

To select an appropriate energy cut-off and k-points grid, several convergence tests were conducted, with the results shown in **Fig. S1** of the supporting information (SI). The plane wave energy cut-off convergence test has been performed over a range of 100 to 1000 eV. As depicted in **Fig. S1**, the total energy converges to a reasonable accuracy beyond an energy cut-off value of 450 eV. Therefore, we chose a cut-off of 600 eV to balance calculation accuracy and computational efficiency. Similarly, the k-points grid was used to divide the first Brillouin zone for energy calculations. A convergence test for the k-mesh was conducted using grid sizes from 1×1×1 to 35×32×1. Although the system's total energy converged at a k-points grid of 12×9×1 (see **Fig. S1**), we opted for a grid size of 25×21×1 to ensure robustness. The proposed crystal structures were fully relaxed using a conjugate-gradient algorithm with a force tolerance of $10^{-3}$ eV/Å per atom. Additionally, the energy tolerance throughout the calculations was maintained at $10^{-8}$ eV. Both structures have rectangular unit cells with periodic conditions in the x-y plane, and a vacuum of 20 Å was applied to avoid interactions along the z-axis. Electronic band calculations were performed for both monolayers along the high symmetry points Γ−X−C−Y−Γ.

We investigated the phonon dispersion to demonstrate the dynamical stability of the bpn-BCN monolayer. For dynamic studies, a supercell is generally required to capture long-wavelength fluctuations in atomic movements within the periodic structure. We used the phonopy package and VASP to calculate the force sets and force constants for the phonon calculations. For phonon calculations and ab initio molecular dynamics (AIMD) simulation, a supercell size of 4×3×1 and a k-mesh of 9×11×1 were used for the I-BPN monolayers. Since the reciprocal cell is compressed due to the supercell in the direct lattice, a less fine k-mesh suffices for k-point sampling. A Nose thermostat NVT canonical ensemble was employed with a time step of 1 fs over 5000 fs at 800K temperature.

To study the mechanical properties, we utilize the generalized Hooke's law, expressed as:

$$\sigma_{ij} = C_{ijkl}\epsilon_{kl}; C_{ijkl} = \frac{1}{2}\frac{\partial^2 U}{\partial \epsilon_{ij} \partial \epsilon_{kl}} (where\, i,j,k,l = 1,2,3) \quad (3)$$

where σ and ϵ represent the second-rank stress and strain tensors, and C is the fourth-rank stiffness constant. These stiffness constants are further denoted as $C_{ijkl} \to C_{pq}$ (p,q=1,2,3,4,5,6) in Voigt notation[41]. For illustration ij→p is given as 11→1, 22→2, 33→3, 21=12→4, 13=31→5, 23=32→6.

The elastic strain energy per unit area can be dented as;

$$U(\epsilon_{11}, \epsilon_{22}) = \frac{1}{2} C_{11}\epsilon_{11}^2 + C_{22}\epsilon_{22}^2 + C_{12}\epsilon_1\epsilon_2 + \cdots \quad (4)$$

Here, $\epsilon_{11} \wedge \epsilon_{22}$ represents the strain along the x and y-axes, and $C_{11}, C_{22}, C_{44} \wedge C_{12}$ are stiffness constants in Voigt notation.

Additionally, Young's modulus $Y(\ )$ and Poisson's ratio $\nu(\theta)$ along any arbitrary in-plane direction (where θ is the angle for the x-direction) are determined using the following formulas:

$$Y(\ ) = \frac{C_{11}C_{22} - C_{12}^2}{C_{11}s^4 + C_{22}c^4 + \left(\frac{C_{11}C_{22} - C_{12}^2}{C_{44}} - 2C_{12}\right)c^2s^2} \quad (5)$$

$$\nu(\ ) = \frac{-\left(C_{11} + C_{22} - \frac{C_{11}C_{22} - C_{12}^2}{C_{44}}\right)c^2s^2 - C_{12}(s^4 + c^4)}{C_{11}s^4 + C_{22}c^4 + \left(\frac{C_{11}C_{22} - C_{12}^2}{C_{44}} - 2C_{12}\right)c^2s^2} \quad (6)$$

where $c = \cos \wedge s = \sin$ based on the calculated elastic constants.

Lattice transport properties, including phonon thermal conductivity ($\kappa_{ph}$), were determined using the Phono3py package.[42] A supercell of 4×3×1 and k-mesh of 9×11×1 has been utilized for the bpn-BCN structure to compute lattice thermal conductivity. The Phonopy[42] package has been employed to derive second-order force constants, employing symmetric displacements to calculate the forces necessary for dynamical matrices. The same pseudopotentials and plane-wave basis cut-off energy were applied, along with a 9×11×1 k-mesh grid. Third-order anharmonic interatomic force constants (IFCs) have been calculated by considering interactions up to three nearest neighbours. This resulted in 1902 displacement datasets for both bpn-BCN structures, with an atomic displacement of 0.01 Å. The second and third-order IFCs were then used as input to integrate the linearized phonon Boltzmann transport equation. A dense q-mesh of size 120×120×1 has been employed for the cumulate of $\kappa_{ph}$.

In single-mode relaxation time approximation (SMRTA) within Phono3py[43, 44], the phonon thermal conductivity can be written in the form;

$$\kappa_\lambda^{\alpha\alpha} = \frac{1}{NV} \sum_\lambda C_\lambda v_\lambda^\alpha \quad {}_\lambda^\alpha = \frac{1}{NV} \sum_\lambda C_\lambda v_\lambda^\alpha \otimes v_\lambda^\alpha \tau_\lambda. \quad (7)$$

Here, $C_\lambda$ is the phonon heat capacity and defined as

$$C_\lambda = k_B \frac{(\hbar\omega_\lambda/k_BT)^2 e^{\hbar\omega_\lambda/k_BT}}{\left(e^{\hbar\omega_\lambda/k_BT} - 1\right)^2} \quad (8)$$

$v_\lambda^\alpha$ signifies the phonon group velocity along α-direction (α = x/y/z) and is defined as $v_\lambda^\alpha = \frac{\partial \omega_\lambda}{\partial q_\alpha}$, λ represents phonon mode as the pair of phonon wave vector $q$ and branch $j$, ${}_\lambda^\alpha$ denotes the projection of the mean free displacement along α-direction and $\tau_\lambda$ is single mode phonon relaxation time (phonon lifetime). The phonon lifetime $\tau_\lambda$, has been derived from the phonon linewidth, $2\Gamma_\lambda(\omega| \ |\lambda)$, associated with the phonon mode $\lambda$.[43]

$$\tau_\lambda = \frac{1}{2\Gamma_\lambda(\omega| \ |\lambda)} \quad (9)$$

The expression for $\Gamma_\lambda(\omega| \ |\lambda)$, is calculated from the Fermi golden rule:

$$\Gamma_\lambda(\omega| \ |\lambda) = \frac{18\pi}{\hbar^2} \sum_{\lambda'\lambda''} |\varphi_{-\lambda\lambda'\lambda''}|^2$$

where $\varphi$ defined the interaction strength between three phonons $\lambda, \lambda', \lambda''$ and $n_\lambda$ is the phonon occupation number. Other parameters $N$, $V$ and $T$ represent the total number of $q$-points (number of unit cells in the real space) in the discretized BZ, volume and temperature of the system, respectively. The $\omega_\lambda = \omega_\lambda(q,j)$ is the phonon angular frequency, $k_B$ denote the Boltzmann constant and $\hbar$ is the reduced Planck constant. The $C_\lambda$ is usually determined using the harmonic approximation, which assumes linear interactions between atoms. However, calculating $\tau_\lambda$ necessitates incorporating anharmonic effects, as this approach considers phonon-phonon interactions that influence scattering processes.

To discuss the phonon mode properties to the phonon thermal conductivity we define a DOS like quantity as

$$\kappa_\lambda^{\alpha\alpha}(\omega) = \frac{1}{N} \sum_\lambda \kappa_\lambda^{\alpha\alpha} \delta(\omega - \omega_\lambda), \quad (11)$$

so that $\kappa_{ph}^{\alpha\alpha}$ to be

$$\kappa_{ph}^{\alpha\alpha} = \int_0^\infty \kappa_\lambda^{\alpha\alpha}(\omega) d\omega \quad (12)$$

where, $\frac{1}{N} \sum_\lambda \delta(\omega - \omega_\lambda)$ is the phonon density of state. The cumulative phonon thermal conductivity is calculated as

$$\kappa_{ph}^{c,\alpha\alpha}(\omega) = \int_0^\omega \kappa_\lambda^{\alpha\alpha}(\omega') d\omega' \quad (13)$$

where $\kappa_\lambda$ denotes the contribution to $\kappa$ for $\lambda$ phonon mode. Further, the mode Grunesian parameters $(\gamma_\lambda)$ is given as

$$\gamma_\lambda = \frac{-V}{\omega_\lambda} \frac{\partial \omega_\lambda}{\partial V} \quad (14)$$

The electronic transport properties have been examined using semi-classical Boltzmann transport equations with energy-independent relaxation time and rigid band approximations facilitated by the BoltzTraP program.[45] These equations enable the expression of various thermoelectric-related variables, such as the electrical conductivity ($\sigma^{\alpha\beta}$), conductivity associated with the thermal gradient, and electronic thermal conductivity $\mathscr{k}_e^{\alpha\beta}$, along the α and β directions.

$$\sigma^{\alpha\beta}(T;\mu) = \frac{1}{\Omega} \int \sigma^{\alpha\beta}(\varepsilon) \left[ \frac{-\partial f_\mu(T;\varepsilon)}{\partial \varepsilon} \right] d\varepsilon \quad (15)$$

$$v^{\alpha\beta}(T;\mu) = \frac{1}{eT\Omega} \int \sigma^{\alpha\beta}(\varepsilon)(\varepsilon - \mu) \left[ \frac{-\partial f_\mu(T;\varepsilon)}{\partial \varepsilon} \right] d\varepsilon \quad (16)$$

$$k_e^{\alpha\beta}(T;\mu) = \frac{1}{e^2\Omega T}\int \sigma^{\alpha\beta}(\varepsilon)(\varepsilon-\mu)^2 \left[\frac{-\partial f_\mu(T;\varepsilon)}{\partial \varepsilon}\right] d\varepsilon \quad (17)$$ while using the above tensor quantities, where T is absolute temperature, $\Omega$ is cell volume, μ is chemical potential and f is Fermi-Dirac distribution, respectively. the Seebeck coefficient ($S^{\alpha\beta}$) can be calculated by the below equation

$$S^{\alpha\beta} = \sum_\gamma \frac{v^{\beta\gamma}}{\sigma^{\alpha\gamma}} \quad (18)$$

$\sigma^{\alpha\beta}(\varepsilon)$ represents the energy-dependent conductivity tensor, which is expressed by

$$\sigma^{\alpha\beta}(\varepsilon) = \frac{1}{N}\sum_{i,k} \sigma^{\alpha\beta}(i,k)\,\delta(\varepsilon - \varepsilon_{i,k}) \quad (19)$$

Where *N* are the number of k-points, $\varepsilon_{i,k}$ are electron-band energies with band index i and $\sigma^{\alpha\beta}(i,k)$ represents the conductivity tensor as given below,

$$\sigma^{\alpha\beta}(i,k) = e^2 \tau_{i,k}\vartheta^\alpha(i,k)\vartheta^\beta(i,k) \quad (20)$$

Here, e and $\tau_{i,k}$ is the charge of the electron and relaxation time, $\vartheta^\alpha(i,k)$ and $\vartheta^\beta(i,k)$ are group velocities represented as, $\vartheta^\alpha(i,k) = \frac{1}{\hbar}\frac{\partial \varepsilon_{i,k}}{\partial k_\alpha}, \vartheta^\beta(i,k) = \frac{1}{\hbar}\frac{\partial \varepsilon_{i,k}}{\partial k_\beta}$ with $\alpha, \beta$ are tensor indices.

Further, the Seebeck coefficient is dependent on effective mass $m^*$ and n number density can be represented as,

$$S = \frac{8\pi^2 k_B^2 T}{3eh^2} m^* \left(\frac{\pi}{3n}\right)^{\frac{2}{3}} \quad (21)$$

The value of relaxation time ($\tau$) must be computed in order to get the absolute value of these coefficients because BoltzTraP integrates electrical and electronic thermal conductivity in terms of $\tau$. We determine $\tau$ by applying the deformation potential (DP) theory by using effective mass ($m^*$) and mobility (mob.) of the charge carriers.[46] Furthermore, the effective mass of the electron ($m_e$) and hole ($m_h$) has been estimated by using the parabolic curvature of the conduction (for electrons) and valence (for holes) band edges close to the Fermi level, respectively. The mathematical expression for $m^*$ is

$$m^* = \frac{\hbar^2}{\frac{\partial^2 E}{\partial k^2}} \quad (22)$$

The formula for isotropic carrier mobility in 2D systems is given by:

$$mob_{2D} = \frac{e\hbar^3 C_{2D}}{K_B T m^* m_d E_P^2} \quad (23)$$

Whereas, **Eq. 22** overestimates $mob_{2D}$ for anisotropic 2D materials.[47] Therefore, the correction for anisotropy has been incorporated into **Eq. 23**.[48]

$$mob_x = \frac{e\hbar^3 \left(\frac{5C_x + 3C_y}{8}\right)}{K_B T m_x^{3/2} m_y^{1/2} \left(\frac{9E_x^2 + 7E_x E_y + 4E_y^2}{20}\right)} \wedge mob_y = \frac{e\hbar^3 \left(\frac{5C_y + 3C_x}{8}\right)}{K_B T m_y^{3/2} m_x^{1/2} \left(\frac{9E_y^2 + 7E_y E_x + 4E_x^2}{20}\right)} \quad (24)$$

Here, $m_x$ and $m_y$ denotes the effective mass representing armchair(x) and zigzag(y) directions, respectively, $k_B$ is the Boltzmann constant, $E_P$ stands for the deformation potential constant and $C_x$ and $C_y$ represents the elastic moduli ($C_{x/y} = \frac{1}{A_0}\frac{\partial^2 E}{\partial \chi^2}$, where $E, A_0$ and $\chi$ represents total energy in different deformation states, lattice area at the equilibrium and the applied uniaxial strain along the x- and y-axis, respectively) which is determined by quadratically fitting the energy-strain data and $E_{x/y}$ is the DP constant that reflects the strain-induced shift of the band edges (valence band maximum (VBM) for holes and conduction band minimum (CBM) for electrons).

Moreover, the carrier mobility and relaxation time can be calculated using the following relations,

$$\tau = \frac{m}{e} mob_{2D} \quad (25)$$

3. **Results and Discussions**
   3.1. **Crystal structure and its stability**

The optimised bpn-BCN monolayer adopts an oblique Bravais lattice configuration with the space group notation of Pm(6). The optimized lattice parameters have been found to be a = 3.86 Å, b = 4.54 Å, and gamma = 92.5°. The atomic arrangement depicted in **Fig.1(a)** represents the most stable isoelectronic phase of B(green), C(brown), and N(silver) within the biphenylene carbon network. Our previous work extensively investigated various isoelectronic phases of BCN in biphenylene. This specific atomic arrangement proved the most feasible because of their better structural stability, mechanical strength and formation energy.

The formation energy of the $E_{form}$ monolayer is calculated using the equation:

$$E_{form} = \frac{E_{bpn-BCN} - (2\mu_B + 2\mu_C + 2\mu_N)}{6} \quad (26)$$

Here, $E_{bpn-BCN}$ is the total energy of the bpn-BCN monolayer. The chemical potentials are as follows: $\mu_B$ is the chemical potential of boron in a B-rich environment, taken from the total energy per atom of the metallic alpha-B phase; $\mu_C$ is the chemical potential of carbon from graphite; and $\mu_N$ is the chemical potential of nitrogen in an N-rich environment, derived from the total energy per atom of the alpha-N$_2$ phase of solid nitrogen. These values are reported in our previous work.[49] The $E_{form}$ of bpn-BCN is -0.16 eV per atom, which is smaller than that of BPN (-0.54 eV) and I-BPN (-0.35 eV).[18] However, the negative value indicates that bpn-BCN synthesized process can be an exothermic process. This stability primarily stems from the maximised numbers of C-C and B-N bonds per unit cell. The B-C and C-N bonds are comparatively weaker than C-C and B-N in different coordination numbers. Within this structure, octagonal, hexagonal, and square motifs emerge, showcasing a hybridised arrangement of B, C, and N atoms. The electron localisation function (ELF) illustrated in **Fig.1(b)** offers valuable insights into the bonding nature of the BCN monolayer. In this visualization, fully localized electron regions are highlighted in red (1), half-delocalized electron regions in green (0.5), and areas with low electron density in blue (0). The ELF analysis reveals distinct characteristics: the C-C and B-C bonds exhibit clear covalent bonding because the ELF value is high at the centre of the bond, whereas, in the B-N and C-N bonds, there is a slight polarization of electrons towards the N atoms, indicating a partially ionic nature in these bonds.

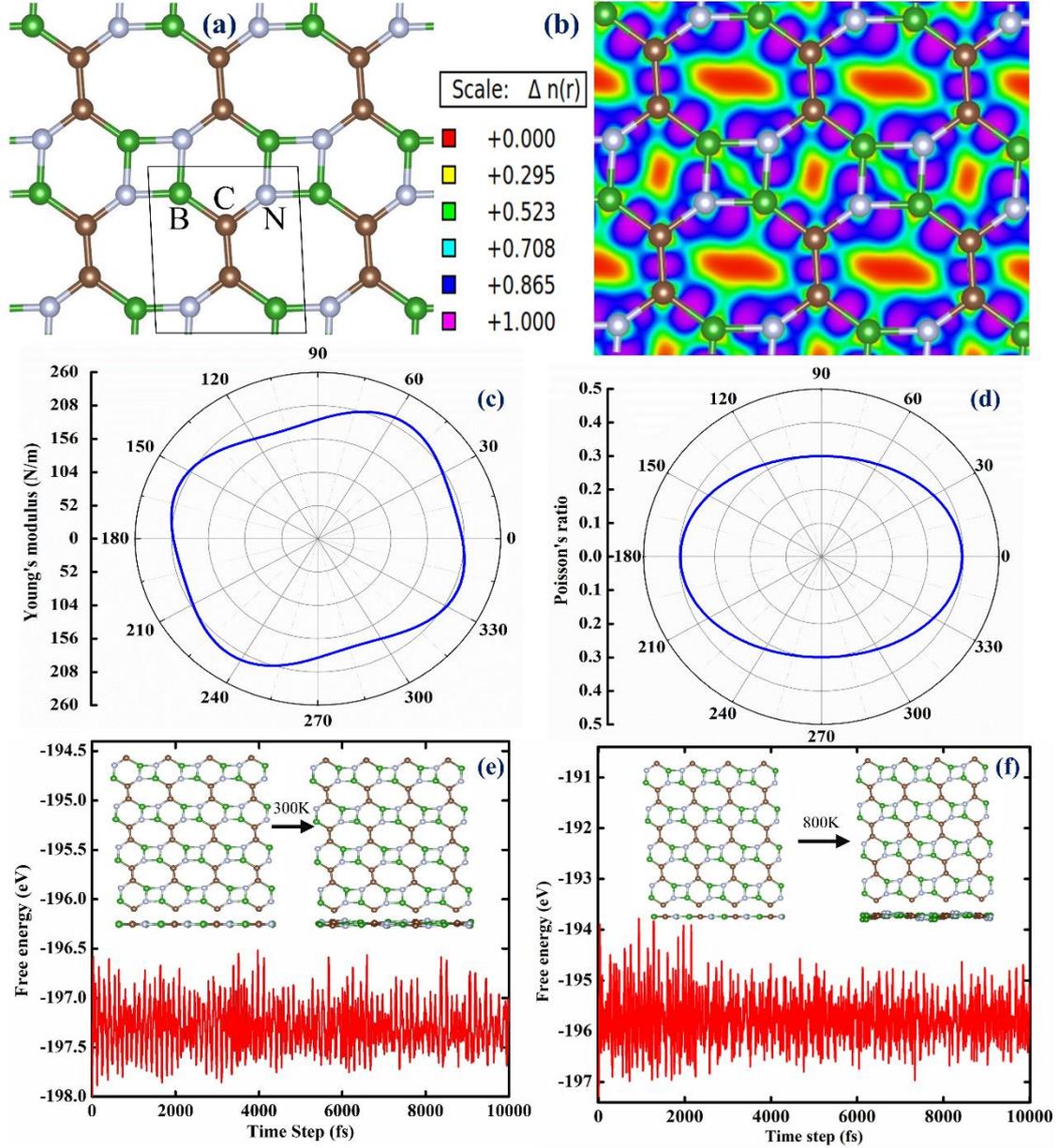

*Fig.1:* *(a) The crystal structure of bpn-BCN monolayer, (b) electron localised function, (c) and (d) direction depending Young's modulus and Poisson ratio (e) and (f) AIMD at 300K and 800K with a time step of 1fs over the 10 ps period.*

The mechanical stability of the bpn-BCN monolayer has been assessed through Born–Huang's criteria,[50] utilizing elastic constants $C_{ij}$ which has been calculated by using **Eq. 3** and **4**. The calculated $C_{11}$, $C_{12}$, $C_{16}$, $C_{22}$, $C_{26}$ and $C_{66}$ has been reported in **Table 1**. The elastic coefficients met Born–Huang's criteria, $C_{11}>0$, $C_{12}>0$, $C_{26}>0$ and $C_{11}C_{22}>C_{12}^2$ for the bpn-BCN monolayer, confirming its mechanical stability. Furthermore, the direction depends on Young's modulus Y(θ) and Poisson's ratio ν(θ) of the bpn-BCN monolayer has also been calculated by using **Eq. 5** and **6**. The Y(θ) and ν(θ) values are highly anisotropic in nature, which can be seen in the polar plot depicted in **Fig.1(c)** and **1(d)**. The maximum(minimum) values with Y(θ) and ν(θ) of 219.53(175.68) Nm$^{-1}$ and 0.44(32) with anisotropy of 1.25 and 1.37, respectively. The BCN monolayer exhibits moderate resistance to deformation under compression or tension compared to other BCN family members such as h-BN ( 279.2 Nm$^{-1}$)[51], g-BCN,[52] t-BCN,[53] and O-BCN,[54] which have Y values ranging from 118 to 300 Nm$^{-1}$. The Poisson's ratio range of 0.295–0.34 indicates good directional dependence, making the layered BCN less prone to fracture. Its Poisson ratio is very similar to that of I-BPN and BPN ($v$ = 0.290–0.296) within the biphenylene family.[18, 55] Generally, the bpn-BCN monolayer exhibits lower mechanical stability than the carbon-based biphenylene network. Still, it is more stable than the BN biphenylene network. **Table 1** reported the comparison of mechanical parameters of BPN, I-BPN and bpn-BCN monolayer.

*Table 1: Elastic constants, Young's moduli and Poisson ratio of BPN, I-BPN and bpn-BCN monolayers*

|  | Elastic coefficients (Nm$^{-1}$) | | | | | | Y($\theta$) (Nm$^{-1}$) | | $v(\theta)$ | | Ref. |
| --- | --- | --- | --- | --- | --- | --- | --- | --- | --- | --- | --- |
|  | $C_{11}$ | $C_{12}$ | $C_{16}$ | $C_{22}$ | $C_{26}$ | $C_{66}$ | Max | Min | Max | Min |  |
| BPN | 292.99 | 94.36 | - | 242.63 | - | 83.42 | 256.3 | 212.24 | 0.39 | 0.32 | 18, 55 |
| I-BPN | 219.30 | 70.34 | - | 191.82 | - | 169.2 | 193.50 | 169.25 | 0.36 | 0.32 | 18 |
| bpn-BCN | 242.14 | 89.51 | 1.41 | 220.94 | 12.3 | 76.51 | 219.53 | 175.68 | 0.44 | 0.32 | This work |

The thermal stability of the bpn-BCN has been checked through ab initio molecular dynamics (AIMD), simulating its response to heating to 300 K and 800 K for 10 ps. **Fig.1(e)** and **1(f)** illustrate minimal fluctuations in the total energy of the structure during the initial temperature rise, suggesting resilience to extreme thermal conditions, especially at 800K. Despite the initial perturbations, the bpn-BCN structure quickly reverts to near-equilibrium states, leading to gradual fluctuations in both total energy and temperature over time. These observations, depicted in **Fig.1(e-f),** indicate the robust thermal stability of the layered bpn-BCN, with only slight alterations in its atomic structure throughout the AIMD simulation.

The details of the possibility of synthesis of bpn-BCN have been discussed in our previous study.[49] In brief, three main precursors can be used: 1,2-BN cyclohexane,[56] bis-BN cyclohexane,[57] and 1,6;2,3-bis-BN cyclohexane[58] can be seen in **Fig. S2**. Dai et al. recently synthesized 1,6;2,3-bis-BN cyclohexane. The dehydrogenated precursor molecules act as free radicals and interact with other molecules, forming a 2D network. The interaction location of either 1,2-BN cyclohexane, 1,6;2,3-bis-BN cyclohexane, or both combined could determine the formation of all phases of bpn-BCN in thin film deposition.

### 3.2. **Phonon transport properties**

**Fig.2(a)** illustrates the phonon dispersion of bpn-BCN along high-symmetry directions, calculated with second-order (harmonic) force constants obtained from first-principles calculations. The absence of imaginary lines in the phonon band structure confirms the thermodynamic stability of the bpn-BCN monolayer. The high symmetry points, as depicted in **Fig.2(a),** are determined based on the unit cell of the bpn-BCN monolayer. With six atoms in the primitive cell, there are eighteen branches in the phonon dispersion. At the $\Gamma$-point, the three lowest phonon branches represent the acoustic modes: ZA (out-of-plane flexural acoustic), TA (in-plane transverse acoustic), and LA (in-plane longitudinal acoustic), as labelled in **Fig.2(a)**. The TA and LA modes exhibit linear behaviour at the $\Gamma$-point, while the ZA branch displays a slightly quadratic nature near the $\Gamma$-point due to rotational symmetry. The highest frequency, located along the H high symmetry point, is 44.86 THz, slightly lower than that of graphene.

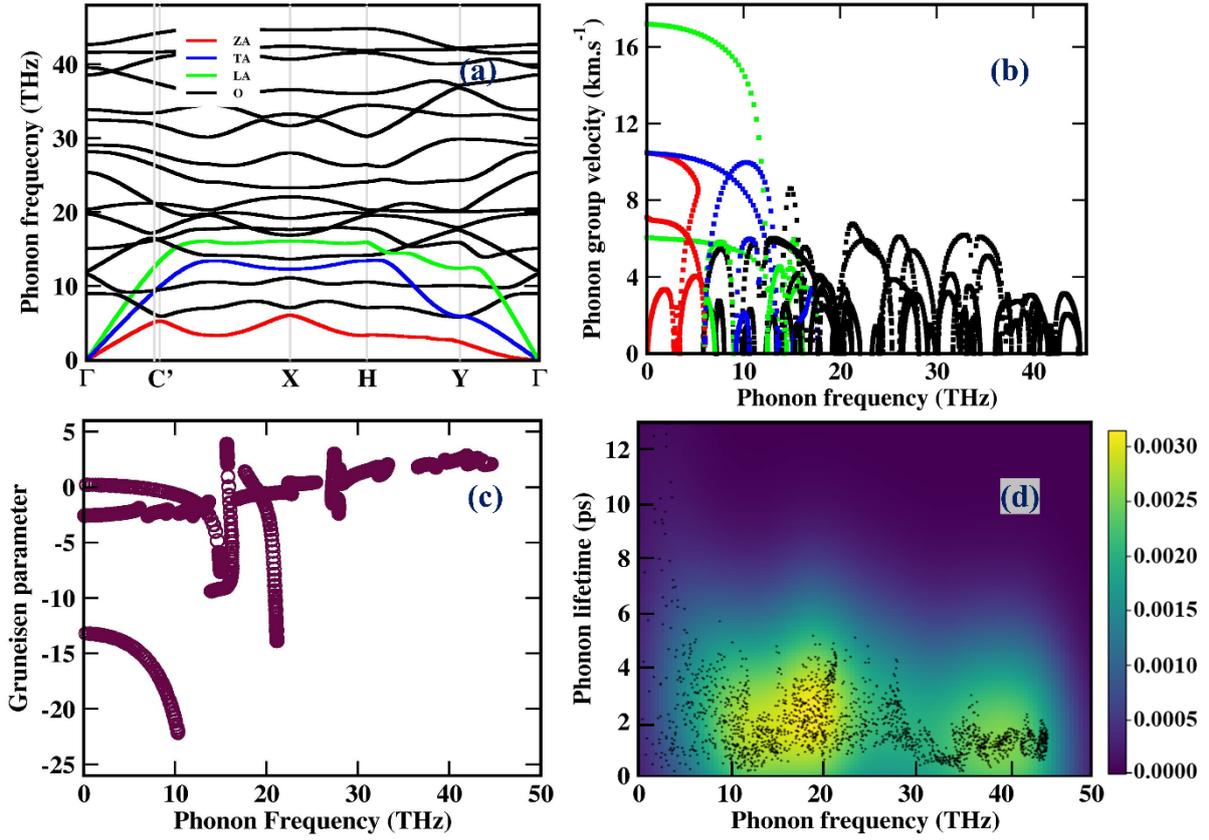

***Fig.2:*** *(a) Phonon dispersion spectrum, (b) phonon group velocity, (c) Gruneisen parameter, (d) phonon lifetime of the bpn-BCN monolayer with colormap represents the density ratio.*

Phonon group velocities have been calculated to understand the thermal transport properties of the bpn-BCN monolayer derived from the slope of the phonon dispersion shown in **Fig. 2(b)**. These velocities are crucial for understanding thermal transport in the material. The phonon group velocity in the ZA mode is small near the Γ-point as it tends to the quadratic nature of the phonon dispersion. In contrast, the LA mode exhibits the largest phonon group velocities, reaching 17.8 km.s$^{-1}$ along the Γ–C'. The group velocity of optical phonon bands is much smaller than that of acoustic modes signifies that the major contribution of lattice thermal conductivity is from acoustic modes. This variation in phonon group velocities is attributed to the anisotropic bond strength, as the phonon group velocity is directly proportional to the bond strength.

The Gruneisen parameter (γ) reflects phonon anharmonicity in a material, which is crucial for understanding lattice thermal conductivity. **Fig. 2(c)** illustrates the correlation between the Gruneisen parameter and phonon frequency and depicts how changes in crystal volume reflect the degree of lattice anharmonicity. The volume-dependent mode γ varies widely across positive and negative values, averaging -1.46 at room temperature (T = 300 K). This average of Gruneisen parameter γ=-1.46 indicates significant anharmonicity, correlating with lattice thermal conductivity. Further, lattice anharmonicity is attributed to the flexibility of the C-C, B-N, and C-N bonds in the x-y plane. Notably, the contribution to the mode γ primarily stems from acoustic phonon modes and overlapping acoustic and optical modes. The small ratio of carbon-to-boron and nitrogen masses often results in the significant overlap between their acoustic and optical phonons, leading to a clustering of the acoustic phonon dispersions. This overlapping greatly influences the intrinsic thermal conductivity, with the LA mode playing a pivotal role in thermal transport along the a-axis. In contrast, the TA mode predominantly governs thermal transport along the b-axis[21]. The negative value of γ in **Fig. 2(c)** indicates an unconventional phenomenon known as negative thermal expansion behaviour in bpn-BCN. Acoustic phonon modes, especially in the low-frequency range, exhibit higher negative γ values than optical modes. Anharmonicity determines the strength of the three-phonon scattering process, with higher anharmonicity leading to stronger phonon–phonon interactions, resulting in lower phonon lifetimes and thermal conductivity. Considering three-phonon interactions, the phonon lifetimes are calculated and plotted in **Fig. 2(d)**. The phonon lifetime is inversely proportional to the square of the frequency ($\tau_{ph} \propto 1/\omega^2$), which aligns well with Klemens's prediction.[59] Additionally, the phonon lifetimes of the acoustic branches are significantly higher than those of the optical branches, indicating longer phonon lifetimes and suggesting a major contribution to the overall thermal conductivity.

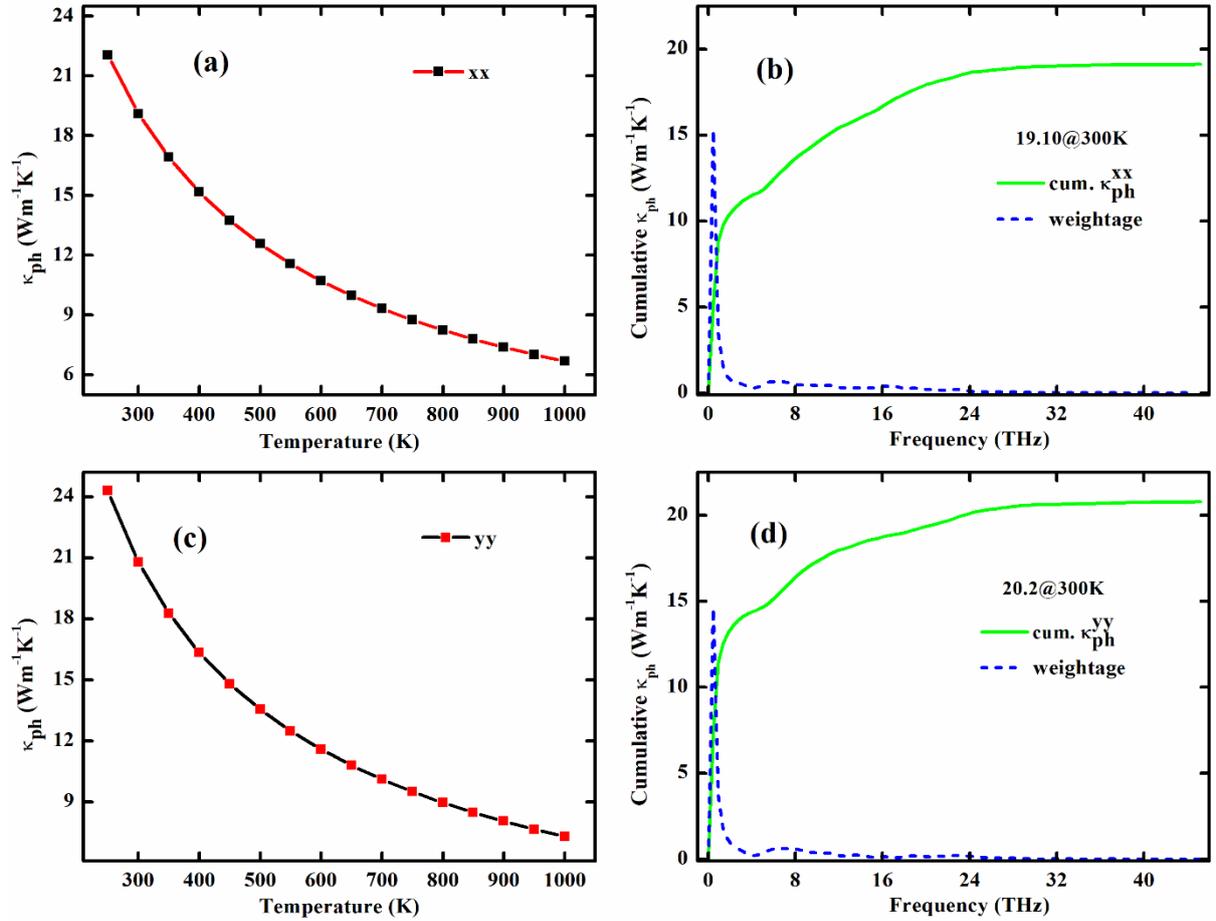

*Fig. 3: (a) and (c) calculated anisotropic phonon thermal conductivity at different temperatures and cumulative lattice thermal conductivities of (b) and (d) bpn-BCN monolayer.*

The temperature-dependent anisotropic lattice thermal conductivities of the bpn-BCN monolayer are shown in **Fig**. **3(a)** and **(c).** At 300 K, the lattice thermal conductivity along the x-axis and y-axis is found to be 19.10 Wm$^{-1}$K$^{-1}$ and 21.23 Wm$^{-1}$K$^{-1}$, respectively, which is significantly lower compared to the parent carbon-based biphenylene 2D material. To provide a systematic comparison, the lattice thermal conductivities of other similar 2D materials are summarized in **Table 2**. The results also show that Umklapp scattering is dominant in these systems, as evidenced by the 1/T relation indicated in **Fig**. **3**. Additionally, **Fig**. **3(b)** and **(d)** depict the cumulative lattice thermal conductivity, emphasizing the contributions of different frequency modes to the heat transport. Acoustic phonons are found to dominate, contributing 87.16% to the lattice thermal conductivity, while optical phonons account for the remaining 12.84% at 300 K. Further, for optimal thermoelectric materials, both heat capacity and phonon lifetime should be minimized. Generally, heat capacity increases with temperature and eventually saturates, as seen in **Fig. S3 (b)**, whereas phonon lifetime decreases with temperature (see **Fig. S4**), contributing to a reduction in $\kappa_{ph}$. For bpn-BCN monolayer, the maximum value of specific heat at 1000K is around 129.34 JK$^{-1}$mol$^{-1}$ and 62.33 JK$^{-1}$mol$^{-1}$ at room temperature. Additionally, as shown in **Fig. S4**, phonon lifetime decreases with increasing temperature. Interestingly, although both heat capacity and phonon lifetime are proportional to $\kappa_{ph}$, the overall reduction in $\kappa_{ph}$ indicates that the phonon lifetime plays a more dominant role with increasing temperature as heat capacity tends to saturate.

***Table 2*** *The $\kappa_{ph}$ of various 2D materials similar to bpn-BCN monolayer at room temperature.*

| 2D material | $\kappa_{ph}$ $(Wm^{-1}K^{-1}$ | Reference |
|---|---|---|
| Graphene | ~3000-3800 | 60-62 |
| h-BN | 450-600 | 63, 64 |
| h-BCN | ~2-50 | 65, 66 |
| BPN | ~187-310 | 67, 68 |
| I-BPN | ~20-35 | 18 |
| bpn-BCN | ~19-22 | this work |

### 3.3. Electronic bands structure

Since the GGA-PBE approximation often underestimates the bandgap, we employed the more accurate yet computationally demanding HSE06 method for band structure calculations. The electronic properties of the bpn-BCN monolayer were examined by calculating its band structure and projected density of states (PDOS) using the HSE method, as depicted in **Fig. 4**. The unit cell of bpn-BCN, along with its first Brillouin zone and the high-symmetry path, is illustrated in Fig. S6. The atom-projected electronic band structure and DOS highlight the atomic contributions to the valence band maximum (VBM) and conduction band minimum (CBM). The VBM is primarily contributed by N atoms, while the CBM is dominated by C atoms. The band structure reveals a nearly linear dispersion for both the valence and conduction bands, with a direct bandgap of 0.58 eV at the X-point, compared to the GGA-PBE results, which show an indirect bandgap of 0.19 eV, as depicted in **Fig. S5**. In the GGA-PBE calculation, the VBM is located at the Y-point, whereas the CBM is at the X-point. Interestingly, a transition from an indirect to a direct bandgap occurs with a minimal energy difference, resulting in a direct bandgap of 0.22 eV at the X-point. This transition is characterized by a graphene-like linear energy dispersion between the valence and conduction band edges, as illustrated in **Fig. S7**. The VBM and CBM wavefunction plots are shown in **Fig. S7(a)**. For the CBM, the charge is delocalized over the C-B bond, whereas for the VBM, there is a small charge delocalization on the C-C bond with a localized wavefunction on the N atom. Moreover, the VBM and CBM wavefunctions for the HSE bandgap are shown in **Fig. S7(b)**. While the VBM wavefunction does not change significantly, the CBM becomes localized at the N site, leading to an insulating nature and an increased bandgap in the HSE functional. In the HSE calculation, a large DOS peak appears around the CBM, which is a consequence of the flat dispersion of the conduction band as it travels from the X point toward the C and H points in the band structure. Additionally, a larger DOS is found at the VB edge compared to the CB edge, attributed to a small energy level difference between the X and Y points.

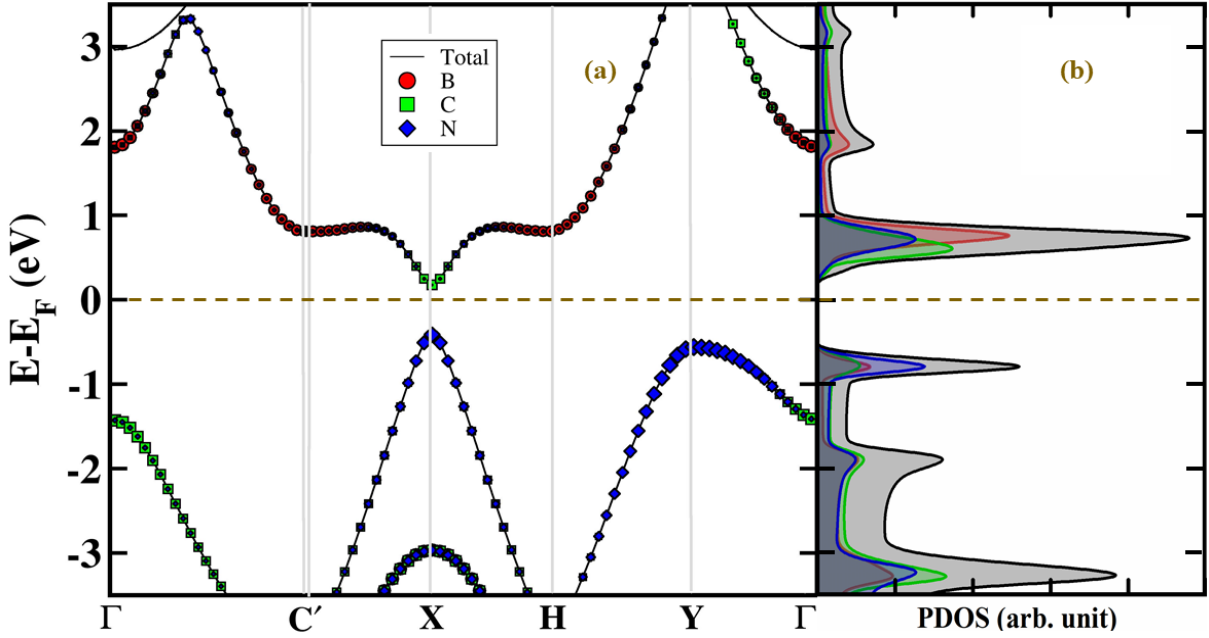

***Fig. 4:*** *(a) Projected electronic band spectra and (b) density of states of the bpn-BCN monolayer.*

### 3.4. Carrier effective mass and mobility

To verify the prediction of high carrier group velocity accompanying linear electronic dispersion, the charge carrier group velocity ($v_g^c$) of bpn-BCN has been calculated with ideal linear band-edge dispersion using the definition where $v_g^c = \frac{dE}{\hbar dk}$ represents the slope of the valence or conduction band edge, and $\hbar$ is the reduced Planck's constant. The calculated group velocities are $4.1 \times 10^5 ms^{-1}$ for electrons and $4.86 \times 10^5 ms^{-1}$ for holes in the bpn-BCN, comparable to the Fermi velocity of electrons in metals and graphene (approximately $10^6$ $ms^{-1}$). High group velocities facilitate the movement of charge carriers, resulting in high conductivity, which can enhance the overall thermoelectric performance of a material. Additionally, using the DP theory, the electronic transport properties of monolayer bpn-BCN have been further estimated by considering anisotropic carrier mobilities. We also reported the deformation potential constant for the charge, their effective mass (m*) at the CBM and VBM, and their stiffness constant along the armchair and zigzag direction. In **Table 3**, the small carrier effective mass can also be anticipated from the linear band-edge dispersion, which deviates significantly from the ideal parabola due to the linear response near the VBM and CBM. The table represents room temperature carrier mobilities in the order of $10^3\ cm^2V^{-1}s^{-1}$. The relaxation time has also been calculated using **Eq. 25**, yielding values of $10^{-13}$s.

| 2D material | bandgap (eV) | $m\ (m_o)$ | | | | $\mu^{2D} \times 10^3 (cm^2V^{-1}s^{-1})$ | | | | Ref. |
|---|---|---|---|---|---|---|---|---|---|---|
| | | e | | h | | e | | h | | |
| | | x | y | x | y | x | y | x | y | |
| BP | 0.91 | 0.18 | 0.10 | 0.17 | 0.09 | 49.96 | 68.81 | 13.7 | 26.05 | 69 |
| Bas | 0.46 | 0.15 | 0.08 | 0.15 | 0.08 | 39.28 | 57.78 | 16.92 | 32.46 | 70 |
| BCP | 0.55 | 0.48 | 0.43 | 0.63 | 0.56 | 1.58 | 2.93 | 1.60 | 0.83 | 71 |
| BCN | 1.9 | 1.22 | 1.61 | 0.94 | 1.24 | 1.76 | 0.88 | 0.23 | 1.44 | 71 |
| I-BPN | 1.88 | 0.74 | 0.51 | 0.70 | 0.67 | 0.44 | 0.26 | 0.34 | 0.18 | 18 |
| bpn-BCN | 0.58 | 0.09 | 0.06 | 0.06 | 0.04 | 0.66 | 0.49 | 0.60 | 0.70 | this work |

*Table 3: Bandgap, effective mass, and charge carrier mobility of 2D materials with similar compositions to bpn-BCN at 300 K.*

Other calculated quantities, like, in-plane stiffness ($C_{2D}$), deformation potential constant ($E_P$) and relaxation time ($\tau$) are presented in **Table S1**. Notably, carrier mobilities exhibit anisotropy, while hole mobilities display strong anisotropy, consistent with prior theoretical findings. Remarkably, a high hole mobility $0.70 \times 10^3 \, cm^2 V^{-1} s^{-1}$ along the armchair direction at room temperature is observed, surpassing the value reported for 2D materials MoS$_2$ ($0.20 \times 10^3 cm^2 V^{-1} s^{-1}$)[72], graphitic-BCN monolayer ($0.176 \times 10^3 cm^2 V^{-1} s^{-1}$).[71] There are other 2D materials like hexagonal BP, BAs, BCP and BCN having the mobilities of the order of $10^3 cm^2 V^{-1} s^{-1}$ as reported in the **Table 3.** The comparison indicates that the bpn-BCN structure has slightly higher carrier mobility than its I-BPN counterpart. This suggests that bpn-BCN has enhanced transport properties compared to its inorganic variant.[18] This enhanced mobility in monolayer bpn-BCN is attributed to its ideal band gap, which facilitates efficient electrical transport. These high carrier group velocities and high carrier mobility clearly indicate a high conductivity with an optimal carrier concentration, making bpn-BCN promising for thermoelectric applications.

### 3.5. Electrical Transport Properties

The Boltzmann transport equation, employed through BoltzTraP software, was utilized to compute the electrical transport properties. This code is well studied and verified with the experimentally reported thermoelectric material such as CoSbS,[73] Bi$_2$Te$_3$.[74] The electronic transport parameters have been calculated by employing an approximation of the rigid band approximation and constant relaxation time. A constant value ($10^{-13} s$) of RT has been used through the calculations which has been estimated by DP theory. **Fig. 5** displays the thermoelectric parameters of bpn-BCN across a chemical potential range of -0.5 to 0.5 eV. The parameters are plotted with positive chemical potentials corresponding to n-type (electron-dominated) behaviour and negative chemical potentials indicating p-type (hole-dominated) behaviour. These parameters are also represented with a temperature range of 200–1000 K at a carrier concentration of $7.89 \times 10^{14} cm^{-2}$ in the **Fig. S8**. For p-type bpn-BCN, the magnitude of the maximum Seebeck coefficient is around $700 \mu V K^{-1}$ at 200 K and $211 \mu V K^{-1}$ at 680 K along the x-axis and y-axis, respectively, as shown in **Fig. 5(a).** Meanwhile, n-type bpn-BCN is approximately $588 \mu V K^{-1}$ at 200 K along the x-axis and $196 \mu V K^{-1}$ at 850 K along the y-axis. Both types exhibit high anisotropy. The high Seebeck coefficient values can be attributed to the flat dispersion of the conduction band along the C to Y direction, crossing a sharp pinch line at the X point in the 1st Brillouin zone. This flat band produces a sharp DOS peak, effectively enhancing the Seebeck coefficient. Additionally, the valence band features degenerate states with a very small energy margin of a few meV, contributing to a small spike in DOS, further enhancing the Seebeck coefficient as per **Eq. 18**.

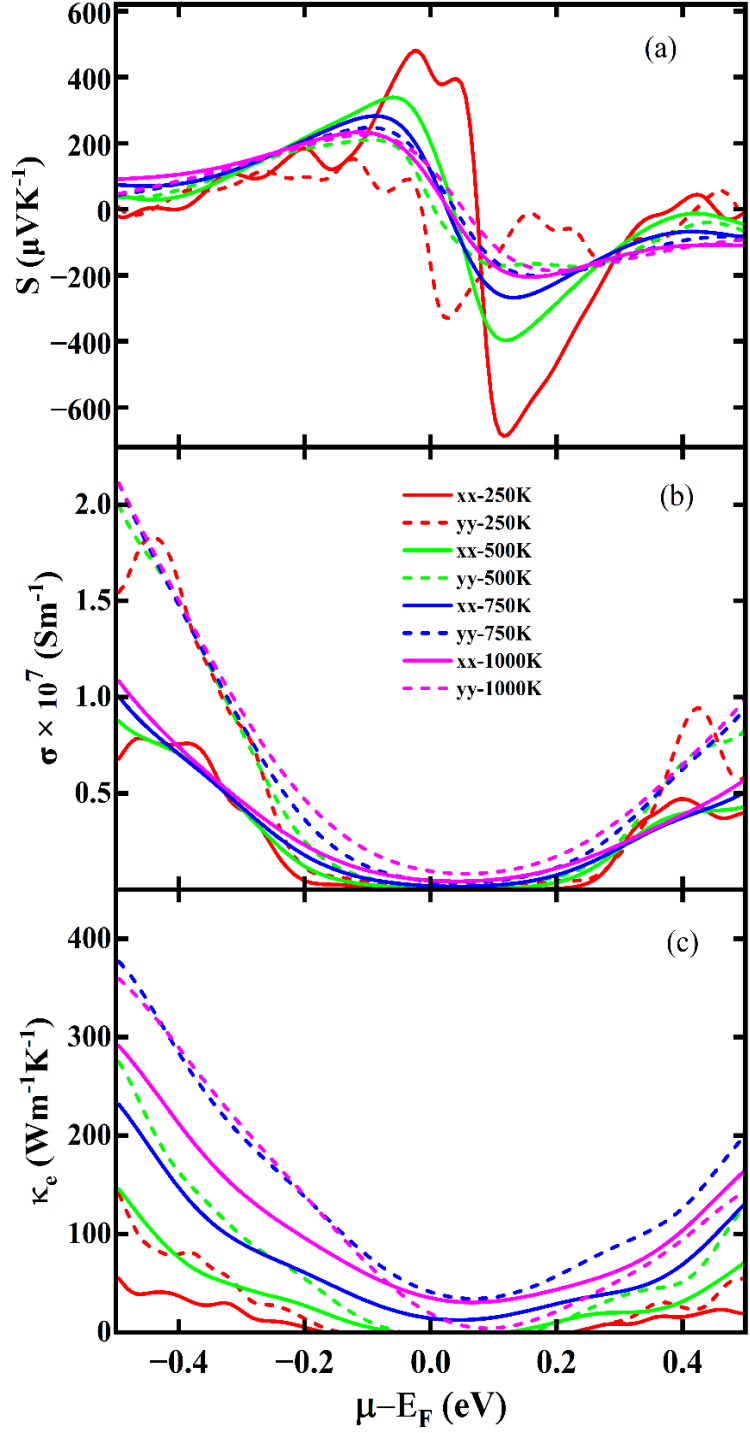

*Fig. 5: Temperature dependence of (a) Seebeck coefficient (b) electrical conductivity (c) are electronic part of thermal conductivity in bpn-BCN monolayer.*

When both electrons and holes contribute to charge transport, significant changes in thermoelectric properties, known as the bipolar effect, occur. This is commonly observed in small band gap materials at higher temperatures. The effect arises when thermal excitation across the band gap generates minority carriers (e.g., holes in n-type and electrons in p-type materials) alongside the majority carriers. The bpn-BCN monolayer exhibits a small direct band gap of 0.58 eV with symmetric band profiles near the CB and VB edge, can be seen in **Fig 4.** As temperature increases, the excitation of minority carriers becomes more prominent due to the narrow band gap, enhancing the contribution of both electron and hole carriers and altering the thermoelectric properties. For p-type doping, as shown in **Fig. S9 (a)**, hole carriers dominate transport, but the contribution of minority electron carriers also affects the Seebeck coefficient (where S changes the sign), especially at low concentration and high temperatures. For instance, at a low temperature of 250 K, the value of S decreases linearly with increasing concentration. However, at higher temperatures such as 750 K and 1000 K, it develops a single-peaked pattern, driven by the enhanced bipolar effect. Similarly, for n-type doping, as shown in **Fig. S9(b)**, where electrons dominate transport, a similar pattern is observed as in the p-type doping case. However, the bipolar effect is not evident at low concentrations at high temperatures. It is possible that the bipolar effect may become noticeable with a further increase in temperature at low concentrations. Additionally, the electrical conductivity increases with temperature as both majority carriers and thermally generated minority carriers contribute to the transport process with the same charge sign. The results presented here confirm that the bpn-BCN structure exhibits the bipolar effect. However, BoltzTraP only partially captures this phenomenon when varying the temperature, as its formulation incorporates the temperature-dependent Fermi-Dirac distribution. The primary limitation lies in the carrier concentration, which is strongly temperature-dependent due to the bipolar effect. In contrast, BoltzTraP calculations are based on the assumption of constant carrier concentration with respect to temperature. This discrepancy makes it challenging to perform self-consistent calculations of thermoelectric parameters without accurate knowledge of the temperature-dependent carrier concentration.[75-77]

**Fig. 5(b)** illustrates the electrical conductivities for bpn-BCN, demonstrating distinct trends. Specifically, the electrical conductivity along the y-axis is higher than along the x-axis for a given doping level. This behaviour is reversed in the Seebeck coefficient, where the x-axis exhibits higher values than the y-axis. Since high electrical conductivity indicates a metallic nature and a high Seebeck coefficient suggests non-conducting properties, these observations reveal that the y-axis is more conductive than the x-axis. This anisotropy highlights the direction-dependent electronic properties of bpn-BCN, with the y-axis favouring conductivity and the x-axis enhancing the thermoelectric voltage generation. At a carrier concentration of $7.8 \times 10^{14} cm^{-2}$, the electrical conductivity of bpn-BCN increases with the temperature for p-type and n-type, are shown in **Fig. S8 (c-d)**. The electrical conductivity can be expressed as $\left(\frac{ne^2\tau}{m}\right)$ where $n$ is the carrier concentration, $e$ is the elementary charge, $\tau$ is the relaxation time, and $m$ is the effective mass. For a constant relaxation time, is inversely proportional to the $m$, as seen in **Fig. 5**. The small $m$ near the VBM and CBM result in high .Additionally, increases with temperature, displaying typical semiconductor behaviour. The higher electrical conductivity of bpn-BCN is due to these smaller effective masses, which enhance mobility. The electronic part of thermal conductivity ($\kappa_e$) also follow the conventional trend, increasing $\kappa_e$ with temperature can be seen in **Fig. 5(c)**. But the order of $\kappa_e$ is smaller than the though both have same energy carrier. This could be because carriers with low effective mass move easily (high mobility), leading to high electrical conductivity. However, their ability to carry heat may be limited compared to their ability to carry charge.

According to Wiedemann-Franz law, $\kappa_e = L\sigma T$, where $L$ is Lorentz number, $\sigma$ is electrical conductivity, and T is temperature. To validate the $L$, the $\sigma$ and $\kappa_e$ at 250K, 500K, 750K and 1000K temperatures have been chosen at a particular value of doping for both n-type and p-type (for instance, μ-E$_F$=0.1 eV). The calculated $L$ values exhibit a slight decreasing trend with temperature, as reported in **Table S2**, which is consistent with the literature.[78, 79]

**Fig. 6(a)** and **6(b)** show the total thermal conductivity $\kappa$ which includes both the phonon $\kappa_{ph}$ and electronic $\kappa_e$ components for n-type and p-type bpn-BCN, exhibiting typical temperature-dependent behaviour for semiconductors in the 200–1000 K range. The $\kappa_{ph}$ of bpn-BCN is anisotropic along the x- and y-axes due to its different atomic environments. At 300 K, the $\kappa_{ph}$ of bpn-BCN is 19.2 $Wm^{-1}K^{-1}$ along the x-axis and 21.4 $Wm^{-1}K^{-1}$ along the y-axis. Although the $\kappa_{ph}$ is very small for bpn-BCN, the total κ is high due to the significant electronic part of thermal conductivity. The low $\kappa_{ph}$ is attributed to a high scattering rate, as indicated by the variation of the phononic lifetime with frequency in the relevant Figure. The average phonon lifetime is 3.72$ps$. at room temperature. Most of the optical modes are involved in three-phonon scattering processes, which impede thermal transport. On the other hand, the acoustic modes exhibit a higher lifetime and significantly contribute to phonon transport. Low $\kappa_{ph}$ is primarily influenced by bond strength and mass symmetry, as well as structural symmetry. In 2D honeycomb structures, C-C bonds exhibit high $\kappa_{ph}$ due to their strong bond strength and mass symmetry, which facilitate efficient phonon transport. Conversely, B-N bonds in h-BN introduce a mass mismatch between boron and nitrogen, increasing phonon scattering and reducing group velocities, resulting in lower $\kappa_{ph}$. When boron, carbon, and nitrogen are combined in a hybrid honeycomb structure, this conductivity is further reduced. Additionally, the BPN, with its hexagons, squares, and octagons, creates spatial irregularities that enhance phonon scattering, resulting in $\kappa_{ph}$ an order of magnitude lower than graphene, despite both being carbon-based. A comparison of $\kappa_{ph}$ between similar 2D materials has been reported in **Table 2**. Further, anisotropic thermal conductivity arises from a hybrid structure of boron nitride and carbon features, where bond strength variations affect phonon velocities. Phonons travel more efficiently along stronger bonding directions and encounter greater scattering in weaker directions. Thus, anisotropy in thermal conductivity along x- and y-axes is influenced by structural symmetry, bond strength disparities, and phonon scattering.

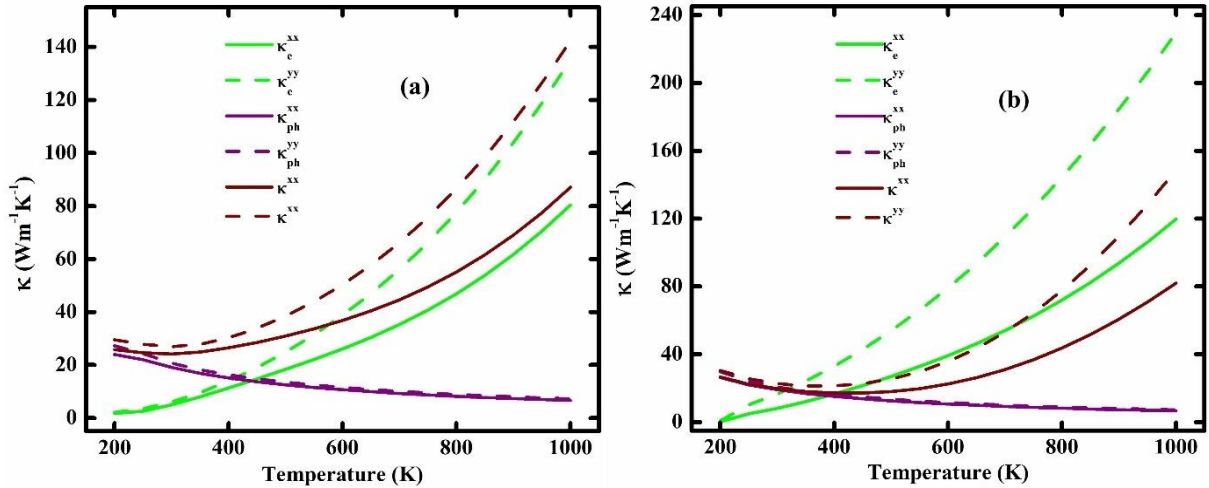

*Fig. 6:* (a) and (b) depict the electronic, phononic and total thermal conductivity of n-type doping and p-type doping bpn-BCN monolayer, respectively.

The power factor, crucial for high-temperature applications, depends on the material's maximum achievable power output and the temperature difference. Enhancing the power factor of bpn-BCN involves increasing either the Seebeck coefficient or the electrical conductivity. However, in bpn-BCN, both the Seebeck coefficient and electrical conductivity are significantly high. The high Seebeck coefficient, stemming from a flat CB and degeneracy of VB and low effective mass, results from linear band dispersion near the VBM and CBM, thereby improving electrical conductivity. The anisotropic nature of power factor varies with temperature for bpn-BCN, as depicted in **Fig. 7(a).** At 1000 K, the maximum PF for the p-type bpn-BCN reaches $93 \times 10^{-3} Wm^{-1}K^{-2}$ along the x-axis and $178 \times 10^{-3} Wm^{-1}K^{-2}$ along the y-axis. For the n-type bpn-BCN, the PF is also anisotropic, with the y-axis exhibiting a significant value of $91 \times 10^{-3} Wm^{-1}K^{-2}$ and the x-axis shows a lower maximum value of $74 \times 10^{-3} Wm^{-1}K^{-2}$. This indicates that the PF along the y-axis is higher than along the x-axis for both p-type and n-type, demonstrating substantial PF anisotropy. The dominant PF along the y-axis in the n-type bpn-BCN monolayer highlights the direction-dependent thermoelectric efficiency of the material.

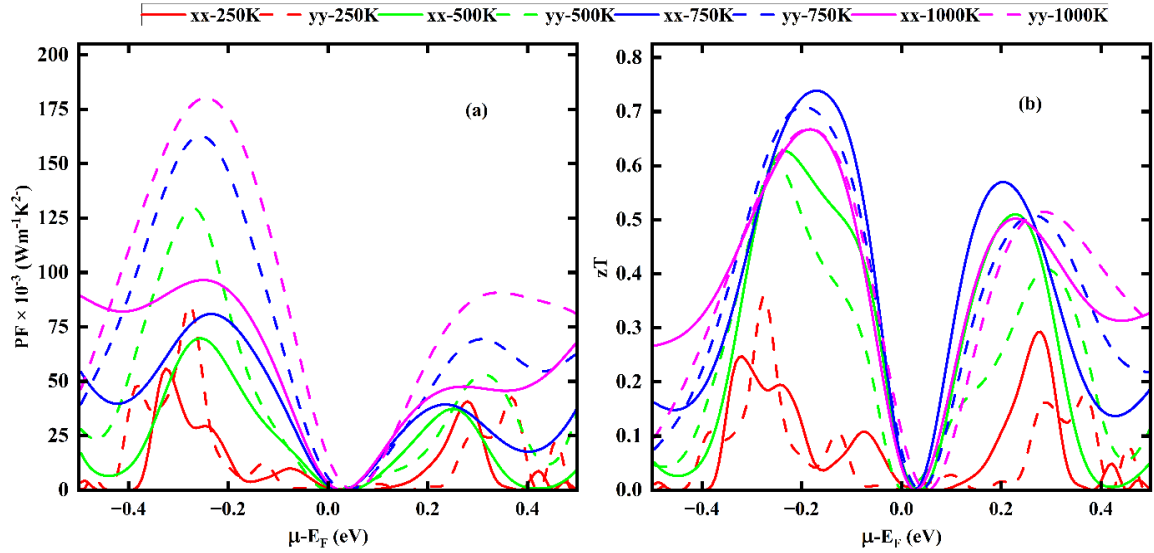

*Fig. 7: Depict (a) the power factor, and (b) the Figure of merit for the bpn-BCN monolayer.*

The Figure of merit for both n-type and p-type bpn-BCN is calculated by combining electronic and phonon thermal conductivity properties, as depicted in **Fig. 7(b)**. The *zT* peak appears at a maximum value of 0.78 (0.72) for p-type at 750 K along the x (y)-axis, while for n-type, the *zT* peak is smaller at 0.57 (0.53) along the x (y) direction at 750 K. The results suggest that *zT* increases with temperature for p-type, reaching a peak value of 0.81 (680 K) along the x-axis and 0.68 (750 K) along the y-axis. Conversely, for n-type, *zT* peaks at 0.77 (200 K) and decreases with temperature along the x-axis, while it reaches 0.57 (750 K) along the y-axis, as shown in **Fig. S10**. These findings demonstrate the temperature-dependent behaviour of *zT* in bpn-BCN, indicating different optimal temperature regimes for maximizing thermoelectric efficiency in p-type and n-type region.

## Conclusions

In summary, our study utilized first-principles computations to explore the structural, mechanical, electronic, and thermal properties of bpn-BCN. We found the material stable regarding formation energy and phonon dispersion without any imaginary frequencies. Although the Young modulus of bpn-BCN is slightly lower than that of experimentally synthesized BPN, both modulus and Poisson's ratio values indicate mechanical stability. Notably, bpn-BCN exhibits lower lattice thermal conductivity than I-BPN and BPN monolayers. Electronic analysis revealed a significant direct bandgap of 0.58 eV (HSE), correcting the underestimated PBE bandgap of 0.19 eV. The PBE bandgap is indirect, with the VBM at the Y point and the CBM at the X point at different K points. Through deformation potential theory, calculations estimated the carrier mobility for electrons (661.41 $cm^2V^{-1}s^{-1}$) which is small compared to holes (707.90 $cm^2V^{-1}s^{-1}$). In terms of thermoelectric properties, bpn-BCN demonstrated high Seebeck coefficients $700 VK^{-1}$ at 200 K for electron-doped regions and $689 VK^{-1}$ for hole-dominant areas. The peak power factor reached $178 \times 10^{-3} Wm^{-1}K^{-2}$ at 1000 K for p-type doping along the y-axis, surpassing the $91 \times 10^{-3} Wm^{-1}K^{-2}$ observed for n-type. Moreover, the highest observed *zT* values were 0.78 (0.72) along the x (y) direction at 750 K for p-type and 0.57 (0.53) at 750 K along the x (y) axis for n-type. The high power factor, particularly for p-type bpn-BCN, driven by its high Seebeck coefficient and electrical conductivity, underscores its potential for an eco-friendly thermoelectric generator.

### Acknowledgement

AK thanks the University Grants Commission (UGC), New Delhi, Government of India, for financial support through a Senior Research Fellowship (DEC18-512569-ACTIVE). PS acknowledges the DST INSPIRE (IF190005) Government of India, for financial assistance. PP thanks the DST-SERB Government of India for the ECRA project (ECR/2017/003305).

### Data availability

The datasets generated during and/or analyzed during the current study are available from the corresponding author upon reasonable request.

### Code availability

Not applicable

### Conflict of interest

On behalf of all authors, the corresponding author states that there is no conflict of interest.